\documentclass[showpacs,pra]{revtex4}
\usepackage{amssymb}
\begin{document}

\title{Lie random fields}
\author{Peter Morgan}
\email{peter.w.morgan@yale.edu}
\affiliation{Physics Department, Yale University, New Haven, CT 06520, USA.}
\homepage{http://pantheon.yale.edu/~PWM22}

\date{\today}
\begin{abstract}
The algebras of interacting ``Lie random fields'' that were introduced in \textit{J. Math. Phys.}
\textbf{48}, 122302 (2007) are developed further.
The conjecture that the vacuum vector defines a state over a Lie random field algebra is proved.
The difference between Lie random field algebras and quantum field algebras is the triviality of the
field commutator at time-like separation, the field commutator being trivial at space-like separation
in both cases.
Many properties that are usually taken to be specific to quantum theory, such as the superposition of
states, entanglement, quantum fluctuations, and the violation of Bell inequalities, are also properties
of Lie random fields.
\end{abstract}

\maketitle

\setlength{\textheight}{1.02\textheight}

\newcommand\Half{{\frac{1}{2}}}
\newcommand\Intd{{\mathrm{d}}}
\newcommand\CC{{\mathbb{C}}}
\newcommand\NN{{\mathbb{N}}}
\newcommand\RR{{\mathbb{R}}}
\newcommand\ZZ{{\mathbb{Z}}}
\newcommand\kT{{{\mathsf{k_B}} T}}
\def\Remove#1{{\raise 1.2ex\hbox{$\times$}\kern-0.8em \lower 0.35ex\hbox{$#1$}}}
\newcommand\eqdef{{\stackrel{\mathrm{def}}{=}}}
\newcommand\DP{{\diamond\kern-0.65em +}}
\newcommand{\SmallFrac}[2]{{\scriptstyle\frac{\scriptstyle #1}{\scriptstyle#2}}}
\newcommand{\BLow}[1]{{\lower 0.65ex\hbox{${}_{#1}\!\!$}}}
\newcommand{\BraVacuum}{{\left<0\right|}}
\newcommand{\Vacuum}{{\left|0\right>}}
\newcommand{\BraN}[2]{{{}_{#1\!}\!\left<#2\right|}}
\newcommand{\KetN}[2]{{\left|#2\right>_{\!#1}}}
\newcommand{\BraKetN}[4]{{{}_{#1\!}\!\left<#3\right|\left.\!#4\right>_{\!#2}}}
\newcommand{\BB}{{(\hspace{-0.2em}(}}
\newcommand{\EB}{{)\hspace{-0.2em})}}

\section{Introduction}
Lie fields were suggested in the 1960s as an elementary way to construct an interacting quantum field
theory that satisfies the Wightman axioms, which was soon found not to be
feasible\cite{GreenbergA,Lowenstein,GreenbergB,Baumann}.
A Lie field was defined to be an operator-valued distribution that satisfies commutation relations
$[\hat\phi_f,\hat\phi_g]=\omega(g,f)+\hat\phi_{\xi(g,f)}$, where the test functions $f$ and $g$  and
$\xi(f,g)$ are typically taken from a Schwartz space, but for a Wightman field that is a Lie field,
it was proved that $\xi(f,g)$ must be zero\cite{Baumann}.
An interesting algebraic structure, a ``Lie random field''\cite{MorganLF}, is possible, however, when we take
the commutation relations for the field to be trivial, $[\hat\phi_f,\hat\phi_g]=0$, which we construct as a sum
of creation and annihilation operators, $\hat\phi_f=a_{f^*}+a_f^\dagger$, where the algebra of creation and
annihilation operators satisfies the commutation relations
$[a_f,a^\dagger_g]=(f,g)+a_{\xi(g,f)}+a^\dagger_{\xi(f,g)}$, following the pattern of the commutation relations
of the Lie field.
We prove the conjecture introduced in \cite[Appendix B]{MorganLF} --- that the map $A\mapsto\BraVacuum A\Vacuum$
that is defined by the vacuum vector is a state over the algebra if $(f,g)$ is an inner product --- in subsection
\ref{VacuumStatePositivity}.

I have pursued the mathematics of Lie random fields in the hope of two possibilities.
Firstly, that from a study of the differences, similarities, and relationships between random fields and
quantum fields and their measurement theories we might gain greater insight into quantum field theory; and
secondly, that the mathematics might be usefully employed in those situations where physicists have
successfully employed other stochastic methods.
In the latter case, Stochastic Electrodynamics (SED)\cite{delaPena} is interesting because it explicitly
models the effects of quantum fluctuations, which has been found to extend considerably the
range of experiments that can be modeled successfully.
There has been little acceptance of such stochastic methods, however, because they appear ad-hoc relative
to algebraic quantum field approaches, a failing that a Lie random field approach to some extent addresses
by itself being a field theory and algebraic.
Lie random fields, as a particular Hilbert space formalism for constructing random fields, have
almost precisely the same properties of superposition and of entanglement as are found in quantum theory.
Note also that the violation of Bell inequalities by experiment does not rule out Lie random fields as a
class of physical models\cite{MorganBellRF}, and there is a particle interpretation of the Lie random
field formalism that is very closely similar to a particle interpretation of a quantum field
theory\cite[\S IV]{MorganLF}.

A random field is a general structure, an indexed set of random variables, that can be used to model
classical stochastic signals.
It is always the case, indeed it's a commonplace, that measurements in classical signal processing
that determine frequency distributions of a signal at a given time are not possible even for
deterministic signals, so that, for example, Wigner functions are frequently used as time--frequency
distributions\cite{Cohen}.
In the Lie random field context, $\hat\phi_f$ commutes with $\hat\phi_g$ for all test functions $f$ and $g$,
but, as in quantum field theory, nonlocal observables such as a projection operator
$a^\dagger_g\Vacuum\BraVacuum a_g$ or $a^\dagger_{g_1}a^\dagger_{g_2}\Vacuum\BraVacuum a_{g_1}a_{g_2}$
generally do not commute, $\left[a^\dagger_f\Vacuum\BraVacuum a_f,a^\dagger_g\Vacuum\BraVacuum a_g\right]$
is generally non-zero independently of whether $f$ and $g$ have space-like separated supports.
Indeed, for a trivial Lie random field and test functions restricted to positive frequency, observables
such as $a^\dagger_g\Vacuum\BraVacuum a_g$ generate exactly the same algebra as they do in a free quantum
field.

With the introduction of creation and annihilation operators and a vacuum state, Lie random fields introduce
noise in an essential way.
The noise is Gaussian for the vacuum state of a free random field, but in general the noise is
non-Gaussian for the vacuum states of the Lie random fields we discuss here.
If the inner product is Lorentz invariant, the noise is not that of thermal fluctuations (which is not
Lorentz invariant), and may reasonably be called quantum fluctuations\cite{MorganQF} (Lorentz invariant
noise is called ``classical electromagnetic zero-point radiation'' in the SED literature).
Random fields are intended here in a relatively empiricist way: a Lie random field is an effective
$\star$-algebraic mathematics for generating mutually consistent probability densities, correlations,
expected values, \textit{etc.}, that are associated with test functions on space-time.
Note, however, that this approach commits us only to a pragmatic mathematical use of probabilities and
statistics, it does not commit us to a frequentist interpretation of probability.

In Section \ref{LTFmain}, we will discuss the commutation relations
\begin{equation}\label{BaseCommRel}
  [a_X,a^\dagger_Y]=(X;Y)+a_{\xi(Y;X)}+a^\dagger_{\xi(X;Y)},\qquad [a_X,a_Y]=0,
\end{equation}
where a set of objects $X_i$ are elements in a vector space $\mathcal{X}$, without specializing to physical models.
For physics, as in Section \ref{Schwartz}, $\mathcal{X}$ is prototypically taken to be a Schwartz space of
functions and the inner product is Lorentz and translation invariant, allowing us to construct models in
space-time, but the commutation relations (\ref{BaseCommRel}) could be used with other types of inner product,
potentially allowing models to be constructed of causal-probabilistic relationships between subsystems that
are not necessarily spatiotemporal.
An account of how to construct suitable Lie random fields is given in Section \ref{CLRF}.
In Section \ref{TsM}, $\mathcal{X}$ will be taken to be a space of functions on the cotangent manifold of
Minkowski space.

\section{Lie random fields}\label{LTFmain}
We consider the commutation relations (\ref{BaseCommRel}).
At first, we consider only the axioms for associativity and for a $\star$-algebraic structure.
Note that we adopt here an opposite convention from that in \cite{MorganLF}, in that $a_X^\dagger$ is taken to be
complex linear, so that $a_X$ is complex anti-linear; we retain the physics convention that the inner product
$(X;Y)$ is complex anti-linear in $X$ and complex linear in $Y$, a convention that we also use for $\xi(X;Y)$.
We obtain an associative $\star$-algebra if $(X;Y)^*=(Y;X)$ and we impose the Jacobi identity,
\begin{equation}
  [a_X,[a_Y,a^\dagger_Z]]+[a_Y,[a^\dagger_Z,a_X]]+[a^\dagger_Z,[a_X,a_Y]]=0,
\end{equation}
which requires that $[a_X,a^\dagger_{\xi(Y;Z)}]=[a_Y,a^\dagger_{\xi(X;Z)}]$,
so that
\begin{equation}
  (X;\xi(Y;Z))=(Y;\xi(X;Z)), \qquad\xi(\xi(Y;Z);X)=\xi(\xi(X;Z);Y),
      \quad\mathrm{and}\quad \xi(X;\xi(Y;Z))=\xi(Y;\xi(X;Z)).
\end{equation}
These relations allow us to write a nested expression such as $\xi(X_1;\xi(X_2;Y_1))$ unambiguously as
$\xi(X_1,X_2;Y_1)$.
More generally, we define $\xi(X_m;\xi(X_1,...,X_{m-1};Y_1,...,Y_n))=\xi(X_1,...,X_{m-1},X_m;Y_1,...,Y_n)$ and
$\xi(\xi(X_1,...,X_{m-1};Y_1,...,Y_n);X_m)=\xi(Y_1,...,Y_n;X_1,...,X_{m-1},X_m)$, and similarly for the 
inner product $(...;...)$.
$(X_1,...,X_m;Y_1,...,Y_n)$ and $\xi(X_1,...,X_m;Y_1,...,Y_n)$ are both symmetric separately in the anti-linear
and linear arguments before and after the ``;''.
In effect, this construction remembers which of its parameters originated as annihilation operators and which
originated as creation operators and it must preserve the commutativity/symmetry associated with creation and
annihilation operators separately.
Given the inner product $(X_1,...,X_m;Y_1,...,Y_n)$ and the linear functional $\xi(X_1,...,X_m;Y_1,...,Y_n)$,
we can define $\xi(;Y_1,...,Y_n)$ and $\xi(X_1,...,X_m;)$ by requiring that
\begin{eqnarray}\label{XiConsistency}
  (U_1,...,U_j;\xi(;Y_1,...,Y_n),V_1,...,V_k)=(U_1,...,U_j;Y_1,...,Y_n,V_1,...,V_k)\cr
  (U_1,...,U_j;\xi(X_1,...,X_n;),V_1,...,V_k)=(U_1,...,U_j,X_1,...,X_n;V_1,...,V_k)
\end{eqnarray}
for all $U_1,...,U_j$ and $V_1,...,V_k$.
In terms of $(X_1,...,X_m;Y_1,...,Y_n)$ and $\xi(X_1,...,X_m;Y_1,...,Y_n)$ and a set of objects
$f_1,...,f_m,g_1,...,g_n\in\mathcal{X}$, we can express the commutation relations (\ref{BaseCommRel}) as
\begin{eqnarray}\label{FCommRel}
  [a_{\xi(f_1,...,f_j;f_{j+1},...,f_m)},a^\dagger_{\xi(g_1,...,g_k;g_{k+1},...,g_n)}]&=&
         (f_{j+1},...,f_m,g_1,...,g_k;f_1,...,f_j,g_{k+1},...,g_n)\cr
          &&\quad+a_{\xi(f_1,...,f_j,g_{k+1},...,g_n;f_{j+1},...,f_m,g_1,...,g_k)}\cr
          &&\quad+a^\dagger_{\xi(f_{j+1},...,f_m,g_1,...,g_k;f_1,...,f_j,g_{k+1},...,g_n)}.
\end{eqnarray}
The action of $\xi$ and the inner product can be extended to the symmetrized tensor algebra over
$\mathcal{X}$ straightforwardly by linearity, except that the inner product is undefined for the scalar
component of the tensor algebra.

\subsection{Modeling compound systems}
The use of Lie random fields extends our ability to model compound systems.
In the vacuum sector, constructed as usual as a zero eigenvector of all annihilation operators, $a_X\Vacuum=0$,
we can construct a vector $a^\dagger_{\xi(;X_1,X_2)}\Vacuum$ as well as the familiar vector constructed
by the action of two creation operators on a vacuum state, $a^\dagger_{X_1}a^\dagger_{X_2}\Vacuum$.
Neither of these vectors is orthogonal to all one-object vectors $a^\dagger_Y\Vacuum$, but for the
difference, the vector
$\KetN{2}{X_1,X_2}=\left(a^\dagger_{X_1}a^\dagger_{X_2}-a^\dagger_{\xi(;X_1,X_2)}\right)\Vacuum$,
we have $\left<Y\right.\!\KetN{2}{X_1,X_2}=0$, for any $X_1$, $X_2$, and $Y$, so it is these differences
that should be considered to be true two-object vectors.
$\KetN{2}{X_1,X_2}$ has the somewhat strange property of being undetectable by any one-object detector,
precisely because $\left<Y\right.\!\KetN{2}{X_1,X_2}=0$ (of course this is just as much a property of
free fields; practical detectors must be modeled as mixtures or superpositions of ideal one- and many-object
detectors).
Because $a^\dagger_{\xi(;X_1,X_2)}$ is effectively trivial in the free field case, we can understand
$a^\dagger_{\xi(;X_1,X_2)}\Vacuum$ as a representation of the emergent properties of $X_1$ and
$X_2$ taken together that is quite distinct from the familiar vector $a^\dagger_{X_1}a^\dagger_{X_2}\Vacuum$.
Table \ref{IPlabel} shows the inner products for the one- and two-object cases, from which we see the way in which
the inner product $(Y_1,Y_2;X_1,X_2)$ (with multiplicity 2) determines the difference between the free field
and an interacting field, corresponding, for example, to the potential energy that determines the difference
between two free particles and two interacting particles in a Hamiltonian formalism.
\begin{table}[ht]
  \begin{tabular}{|c||c|c|c|c|}
    \hline
    \raisebox{2.5ex}{}\raisebox{-2ex}{}
    $\left<\cdot\right|\left.\!\cdot\right>$ &
         $\KetN{1}{X}$ & $a^\dagger_{X_1}a^\dagger_{X_2}\Vacuum$ &
         $\KetN{1}{\xi(;X_1,X_2)}$ & $\KetN{2}{X_1,X_2}$\\
    \hline\hline
    \raisebox{-1.25ex}{}$\KetN{1}{Y}$ & $(Y;X)$ &
                     $(Y;X_1,X_2)$ & $(Y;X_1,X_2)$ & $0$\\
    \hline
    $a^\dagger_{Y_1}a^\dagger_{Y_2}\Vacuum$ & $(Y_1,Y_2;X)$ &
        $\begin{array}{c}(Y_1;X_1)(Y_2;X_2)+(Y_1;X_2)(Y_2;X_1)\\\quad+3(Y_1,Y_2;X_1,X_2)\end{array}$ &
                     $(Y_1,Y_2;X_1,X_2)$ &
        $\begin{array}{c}(Y_1;X_1)(Y_2;X_2)+(Y_1;X_2)(Y_2;X_1)\\\quad+2(Y_1,Y_2;X_1,X_2)\end{array}$\\
    \hline
    \raisebox{-1.25ex}{}$\KetN{1}{\xi(;Y_1,Y_2)}$ &
                     $(Y_1,Y_2;X)$ & $(Y_1,Y_2;X_1,X_2)$ &
                     $(Y_1,Y_2;X_1,X_2)$ & $0$\\
    \hline
    $\KetN{2}{Y_1,Y_2}$ & $0$ &
        $\begin{array}{c}(Y_1;X_1)(Y_2;X_2)+(Y_1;X_2)(Y_2;X_1)\\\quad+2(Y_1,Y_2;X_1,X_2)\end{array}$ &
                               $0$ &
        $\begin{array}{c}(Y_1;X_1)(Y_2;X_2)+(Y_1;X_2)(Y_2;X_1)\\\quad+2(Y_1,Y_2;X_1,X_2)\end{array}$\\
    \hline
  \end{tabular}\caption{Inner Products of the lowest level vectors in the orthogonal sequence}\label{IPlabel}
\end{table}
We can construct a similar three-object vector, which is orthogonal to all two-object and one-object vectors (so
it is undetectable both by one-object and by two-object detectors),
\begin{equation}
  \KetN{3}{X_1,X_2,X_3}=a^\dagger_{X_3}\KetN{2}{X_1,X_2}-\KetN{2}{X_1,\xi(;X_2,X_3)}-\KetN{2}{X_2,\xi(;X_3,X_1)},
\end{equation}
for which there is an emergent behavior associated with all possible partitions, with contributions
to the inner product $\BraKetN{3}{3}{Y_1,Y_2,Y_3}{X_1,X_2,X_3}$ from all distinct permutations of
$(Y_1;X_1)(Y_2,Y_3;X_2,X_3)$ (with multiplicity 2), as well as with the obvious emergent behavior
associated with $(Y_1,Y_2,Y_3;X_1,X_2,X_3)$ (with multiplicity 12), all of which modify the free field
inner product $\sum_{\sigma\in S_3}(Y_1;X_{\sigma(1)})(Y_2;X_{\sigma(2)})(Y_3;X_{\sigma(3)})$
(with multiplicity 1).
Appendix B of \cite{MorganLF} lists this inner product explicitly, and the following subsection allows
the multiplicities to be computed efficiently.

\subsection{Positive semi-definiteness of the vacuum state}\label{VacuumStatePositivity}
We prove here the conjecture posed in \cite[Appendix B]{MorganLF}, that the linear map
$A\mapsto\left<0\right|A\left|0\right>$ is a state over the Lie random field, provided that $(...;...)$ is
positive semi-definite, that is, $\left<0\right|BB^\dagger\left|0\right> \ge 0$ for every element of the
algebra.
It is enough to prove this for $B$ taken to be a sum of products of annihilation operators.
The importance of this is that a state allows us to construct a Hilbert space using the
Gelfand-Naimark-Segal (GNS) construction\cite[\S III.2.2]{Haag}.

$\KetN{1}{X_1}$, $\KetN{2}{X_1,X_2}$, and $\KetN{3}{X_1,X_2,X_3}$ are the first three of a series of
orthogonal vectors, which can be recursively defined as
\begin{equation}\label{KetNdefinition}
  \KetN{n}{X_1,...,X_n}=a^\dagger_{X_n}\KetN{n-1}{X_1,...,X_{n-1}}-
        \sum_{k=1}^{n-1}\KetN{n-1}{X_1,...,\Remove{X_k},...,X_{n-1},\xi(;X_n,X_k)},
\end{equation}
where $\Remove{X_k}$ denotes that this entry in a list is removed.
Alternatively, we can say that the action of $a_Y^\dagger$ on $\KetN{n}{X_1,...,X_n}$ is
\begin{equation}
  a^\dagger_Y\KetN{n}{X_1,...,X_n}=\KetN{n+1}{X_1,...,X_n,Y}+
        \sum_{k=1}^{n}\KetN{n}{X_1,...,\Remove{X_k},...,X_n,\xi(;Y,X_k)}.
\end{equation}
It is straightforward to prove by induction that these vectors, which were introduced in a different
way in \cite{MorganLF} but not extensively studied, have the following illuminating property,
\begin{eqnarray}\label{aKetN}
  a_{Y}\KetN{n}{X_1,...,X_n}&=&\sum_{k=1}^n \left(a^\dagger_{\xi(Y;X_k)}+(Y;X_k)\right)
                                     \KetN{n-1}{X_1,...,\Remove{X_k},...,X_n}\\
     &=&\sum_{k=1}^n \biggl[\KetN{n}{X_1,...,\Remove{X_k},...,X_n,\xi(Y;X_k)}+
          \sum_{j=1\atop j\not=k}^n\KetN{n-1}{X_1,...,\Remove{X_k},...,\Remove{X_j},...,X_n,\xi(Y;X_j,X_k)}+\cr
     &&\hspace{8em}(Y;X_k)\KetN{n-1}{X_1,...,\Remove{X_k},...,X_n}\biggr],
\end{eqnarray}
because
\begin{eqnarray}
    a_Y\KetN{n}{X_1,...,X_n}&=&\left(a^\dagger_{X_n}a_Y+(Y;X_n)+a_{\xi(X_n;Y)}+a_{\xi(Y;X_n)}^\dagger\right)
                             \KetN{n-1}{X_1,...,X_{n-1}}\cr
          &&\qquad   -\sum_{k=1}^{n-1}a_Y\KetN{n-1}{X_1,...,\Remove{X_k},...,X_{n-1},\xi(;X_n,X_k)}\cr
          &=&\sum_{k=1}^{n-1}\left(a^\dagger_{\xi(Y;X_k)}+(Y;X_k)\right)a_{X_n}^\dagger
                                             \KetN{n-2}{X_1,...,\Remove{X_k},...,X_{n-1}}
                                  +\left(a_{\xi(Y;X_n)}^\dagger+(Y;X_n)\right)\KetN{n-1}{X_1,...,X_{n-1}}\cr
          &&\qquad-\sum_{k=1}^{n-1}\sum_{j=1\atop j\not=k}^{n-1}\left(a^\dagger_{\xi(Y;X_j)}+(Y;X_j)\right)
                             \KetN{n-2}{X_1,...,\Remove{X_k},...,\Remove{X_j},...,X_{n-1},\xi(;X_n,X_k)}\cr
          &=&\sum_{j=1}^n \left(a^\dagger_{\xi(Y;X_j)}+(Y;X_j)\right)\KetN{n-1}{X_1,...,\Remove{X_j},...,X_n},
\end{eqnarray}
with other terms cancelling.
Equation (\ref{aKetN}) for the action of $a_Y$ on $\KetN{n}{X_1,...,X_n}$ contrasts with the apparently
more complex expression for the action of $a_Y$ on the elementary vector
$a^\dagger_{X_1}\cdots a^\dagger_{X_n}\Vacuum$, which generates terms with all numbers of creation
operators equal to and less than $n$,
\begin{eqnarray}\label{product1n}
  a_{Y} a^\dagger_{X_1}\cdots a^\dagger_{X_n}\Vacuum &=&
     \sum_{i=1}^n \left(a^\dagger_{\xi(Y;X_i)+(Y;X_i)}\right)
           a^\dagger_{X_1}\cdots \Remove{a^\dagger_{X_i}}\cdots a^\dagger_{X_n}\Vacuum\cr
   &&\quad +\sum_{i=1}^n\sum_{j=i+1}^n\left(a^\dagger_{\xi(Y;X_i,X_j)}+(Y;X_i,X_j)\right)
              a^\dagger_{X_1}\cdots \Remove{a^\dagger_{X_i}}\cdots \Remove{a^\dagger_{X_j}}\cdots a^\dagger_{X_n}
              \Vacuum\cr
   &&\quad +\ \cdots\ +
     \ \sum_{i=i}^n \left(a^\dagger_{\raisebox{0.4ex}[1.8ex]
                              {$\scriptstyle\xi(Y;X_1,...,\Remove{\scriptstyle X_i},...,X_n)$}}+
                          (Y;X_1,...,\Remove{X_i},...,X_n)\right)
            a^\dagger_{X_i}\Vacuum\cr
   &&\quad +\left( a^\dagger_{\xi(Y;X_1,...,X_n)}+(Y;X_1,...,X_n)\right)\Vacuum\cr
   &&\hspace{-4em}
     =\sum_{\sigma\in S_n}\sum_{k=0}^{n-1}\frac{1}{k!(n-k)!}
            \left( a^\dagger_{\xi(Y;X_{\sigma(1)},...,X_{\sigma(k)})}+(Y;X_{\sigma(1)},...,X_{\sigma(k)})\right)
            \left[\prod_{j=1}^{n-k} a^\dagger_{X_{\sigma(k+j)}}\right]\Vacuum.
\end{eqnarray}

From equation (\ref{aKetN}) we immediately observe that
$\left<0\right|a_{Y_1}\cdots a_{Y_m}\KetN{n}{X_1,...,X_n}=0$ for
$m<n$, so that $\KetN{n}{X_1,...,X_n}$ are mutually orthogonal for distinct $n$, no matter what internal
structure $X_1,...,X_n$ may have (such as $X_i=\xi(...;...)\,$).
Applying equation (\ref{aKetN}) twice to $\BraKetN{n}{n}{Y_1,...,Y_n}{X_1,...,X_n}$, which we note is equal
to $\BraN{n-1}{Y_1,...,Y_{n-1}}a_{Y_n}\KetN{n}{X_1,...,X_n}$ because of Equation (\ref{KetNdefinition}),
we obtain
\begin{eqnarray}\label{RecExpr}
  \BraKetN{n}{n}{Y_1,...,Y_n}{X_1,...,X_n}&=&
     \sum_{j=1}^n\BraKetN{n-1}{n-1}{Y_1,...,Y_{n-1}}{X_1,...,\Remove{X_j},...,X_n}(Y_n;X_j)\cr
     &&+\sum_{j=1}^n\sum_{k=1}^{n-1}\BraKetN{n-1}{n-1}{Y_1,...,\Remove{Y_k},...,Y_{n-1},\xi(X_j;Y_n,Y_k)}
                    {X_1,...,\Remove{X_j},...,X_n},
\end{eqnarray}
so that the inner product is manifestly of the form given in \cite[Appendix B]{MorganLF}, and hence that
it is positive semi-definite.
This is enough to prove that $A\mapsto\left<0\right|A\left|0\right>$ is a state over the Lie random field,
provided that $(...;...)$ is positive semi-definite.
For the multiplicity of the terms that are generated by the inner product, we note from equation
(\ref{RecExpr}) that $(Y_1,...,Y_n;X_1,...,X_n)$ is generated
$K_n=n(n-1)K_{n-1}=n!(n-1)!$ times in the product $\BraKetN{n}{n}{Y_1,...,Y_n}{X_1,...,X_n}$, and for a
partition $n_1+\cdots+n_k=n$, the multiplicity of terms that have $k$ inner product factors is
$K_{n_1,...,n_k}=K_{n_1}K_{n_2,...,n_k}=\prod_{j=1}^k K_{n_j}$.

\section{Explicit construction of a class of Lie random fields}\label{Schwartz}
To use observables $\hat\phi_X=a_{X^*}+a^\dagger_X$ to generate consistent classical multivariate probability
densities, we require that $[\hat\phi_X,\hat\phi_Y]=0$, which requires that $(X^*;Y)=(Y^*,X)$ and
$\xi(X^*;Y)=\xi(Y^*;X)$, for all $X,Y\in\mathcal{X}$.
For a construction in space-time, we can take $\mathcal{X}$ to be a Schwartz space of scalar test functions on
Minkowski space that are smooth and approach zero faster than any polynomial at large distances, and
$f\mapsto f^*$ to be real-space complex conjugation at a point, $f(x)\mapsto f^*(x)$ (so that in
Fourier space, $\tilde f(u)\mapsto\widetilde{f^*}(u)=\tilde f^*(-u)$).
For a translation invariant scalar Lie random field, we found in \cite[\S III]{MorganLF} that 
\begin{eqnarray}\label{ScalarLF1}
   (f;g)&=&\quad\int\ \delta^4(u-v)\quad\tilde G^*(u)\tilde f^*(u)\tilde G(v)\tilde g(v)\Intd^4u\Intd^4v,\cr
   \tilde G(s)\tilde\xi(f;g)(s)&=&\lambda\int
                   \delta^4(s+u-v)\tilde G^*(u)\tilde f^*(u)\tilde G(v)\tilde g(v)\Intd^4u\Intd^4v,
\end{eqnarray}
satisfies all the constraints, where $\tilde G(v)=\tilde G^*(-v)$ is a Lorentz invariant function of the
wave-number that models the causal structure of space-time (it was unstated in \cite{MorganLF} that
$\lambda$ must be real).
The condition $\tilde G(v)=\tilde G^*(-v)$ is incompatible with the positive semi-definite Hamiltonian
spectrum condition that is required by the Wightman axioms, for which in this context $\tilde G(v)$ must be
non-zero only for $v$ forward-pointing and either light-like or time-like (see Appendix \ref{PositiveHamiltonian}
for a discussion).
Constructions of a Lie random field based on a vector field or based on the electromagnetic field are
possible, using a symmetrized pair-wise contraction of all tensor indices\cite[\S V]{MorganLF}.

We will reconsider equation (\ref{ScalarLF1}) for a moment.
Recalling that $\widetilde{h^*}(u)=\tilde h^*(-u)$, $[\hat\phi_{h_1},\hat\phi_{h_2}]$ is zero for all $h_1$
and $h_2$ if 
\begin{equation}\label{GGreq}
  \int\tilde G^*(w)\tilde G(w+s)(e^{ix\cdot(2w+s)}-e^{-ix\cdot(2w+s)})\Intd^4w
\end{equation}
is zero for all $s$ and for all $x$ (in a moment we will reconsider microcausality briefly, for which we
require that the Equation (\ref{GGreq}) is zero for all $s$ and for all space-like $x$, but in general is
non-zero for time-like $x$).
In effect, we have here set $h_1(r)=\delta^4(r-x)$ and $h_2(r)=\delta^4(r+x)$; if the commutator is zero
for all such choices, it will be zero for all test functions.
Substituting $z=2w+s$, we require that
\begin{equation}\label{ReqZ}
  \int\tilde G^*\left(\frac{z-s}{2}\right)\tilde G\left(\frac{z+s}{2}\right)
       (e^{ix\cdot z}-e^{-ix\cdot z})\Intd^4z
\end{equation}
is zero for all $s$ and for all $x$.
This is satisfied for time-like as well as for space-like $x$ if
$\tilde G^*(\frac{z-s}{2})\tilde G(\frac{z+s}{2})$ is invariant under $z\mapsto -z$, yielding a Lie random
field, which is easily seen to be possible.

In a slight amendment to the derivation in \cite{MorganLF}, this nonlinear requirement is satisfied for
functions $\tilde G(v)$ for which $\tilde G(v)=c\tilde G^*(-v)$, $|c|^2=1$, not just for
$\tilde G(v)=\tilde G^*(-v)$.
The value chosen for $c$ effectively modifies interference between positive and negative frequency modes.
This has a significant effect for local observables; for example, for the fourth moment of the probability
density associated with an observable $\hat\phi_f$, for which
$\hat\phi_f^\dagger=\hat\phi_{f^*}=\hat\phi_f$, $f^*=f$, we have
\begin{eqnarray}
  \BraVacuum\hat\phi_f^4\Vacuum&=&(f;f,f,f)+4(f,f;f,f)+(f,f,f;f)+3(f;f)^2\cr
                               &=&(4+c+c^*)(f,f;f,f)+3(f;f)^2,
\end{eqnarray}
so we can adjust the kurtosis of this probability density by a factor of $3$ by adjusting the complex phase
of $c$ (parenthetically, the kurtosis cannot in any case be negative, which is a notable limitation).
The value chosen for $c$ makes no difference, however, for nonlocal observables such as
$a_g^\dagger\Vacuum\BraVacuum a_g$ (which are commonly used in quantum field theory), where
$\tilde g(v)$ is non-zero only for $v$ forward-pointing and either light-like or time-like.

Microcausality requires Equation (\ref{ReqZ}) to be zero only for space-like $x$, which requires invariance
of $\tilde G^*(\frac{z-s}{2})\tilde G(\frac{z+s}{2})$ under \emph{all} space-like reflections
$z\mapsto z'=z-2\frac{z\cdot x}{x\cdot x}x$, $x\cdot x<0$, \emph{without} invariance
under $z\mapsto -z$.
Reflections preserve the inner product, $z'\cdot z'=z\cdot z$, and forward and backward light-cones and mass-shells
are separately invariant under space-like reflections so this is equivalent to invariance of
$\tilde G^*(\frac{z-s}{2})\tilde G(\frac{z+s}{2})$ under maps of $z$ to any other point on the same
connected part of its mass-shell, for all $z$ and $s$.
Baumann\cite{Baumann} proves that this construction is not possible for Wightman fields, for which $s$
and $z\pm s$ are restricted to the forward light-cone, except trivially, $\tilde G(v)=0$, but it is
clear from the geometry alone, without the tools of analysis, that nontrivial microcausality is also not
possible for a creation and annihilation operator ansatz for Lie fields even if we do not constrain the
spectrum of the Hamiltonian.

\section{Constructing Lie random fields}\label{CLRF}
It is necessary to build mathematics and intuition that will allow us to match empirical data with
Lie random field models as painlessly as possible.

For a free field, the fundamental mathematical object is the kernel of the map $X\mapsto (X;X)$.
This map is zero everywhere except on the mass shell, and correspondingly there are fluctuations associated
only with the mass shell.
If we attempt to measure $\hat\phi_f$ using a test function that has no on-mass-shell component, we
obtain a probability distribution with zero mean and zero standard deviation, because the inner product
projects all test functions to the mass shell.
We can say, loosely, that only on-mass-shell components of test functions that are used to model
measurement apparatuses ``resonate'' with states in the Hilbert space that are used to model preparation
apparatuses.

For the free electromagnetic field, we can construct a quantum or random
field\cite[\S V]{MorganLF}\cite[equation (3.27)]{MenikoffSharp}, omitting the restriction to positive
frequency in the random field case,
\begin{eqnarray}\label{EMInnerProduct}
     (E;F)_{EMq} &=& \hbar\int\frac{\Intd^4u}{(2\pi)^4}
             2\pi\delta(u_\alpha u^\alpha)\theta(u_0)
             \tilde E_{\mu\beta}^*(u)u^\mu u^\nu\tilde F_{\nu}^{\ \beta}(u)\hspace{2em}\mathrm{(quantum\ field)},\\
     (E;F)_{EMr} &=& \hbar\int\frac{\Intd^4u}{(2\pi)^4}
             2\pi\delta(u_\alpha u^\alpha)\quad
             \tilde E_{\mu\beta}^*(u)u^\mu u^\nu\tilde F_{\nu}^{\ \beta}(u)\hspace{3.2em}\mathrm{(random\ field)}.
\end{eqnarray}
Repeating the brief discussion of \cite[\S V]{MorganLF}, ``note that $E$ and $F$ are \emph{not}
electromagnetic field tensors, they are bivectors of classical test functions that contribute to a
description of measurement and/or state preparation of the quantized electromagnetic field.
If the fourier transform $\tilde F_{\mu\nu}(u)$ of a test function $F$ has electric and magnetic field
components $(\tilde e_1,\tilde e_2,\tilde e_3)$ and $(\tilde b_1,\tilde b_2,\tilde b_3)$, the integrand
for the inner product $(F;F)_{EMq}$ at $(u_0,0,0,u_0)$ is the positive semi-definite form
$u_0^2\left[(\tilde e_1+\tilde b_2)^2+(\tilde e_2-\tilde b_1)^2\right]$, which
suppresses all except two degrees of freedom of the quantum electromagnetic field at each wave number.''
Because we have not included a term
$*\hspace{-0.35ex}\tilde E_{\mu\beta}^*(u)u^\mu u^\nu\!*\hspace{-0.65ex}\tilde F_{\nu}^{\ \beta}(u)$,
where $*\hspace{-0.12ex}F_{\mu\nu}=\varepsilon_{\mu\nu}^{\hspace{1.0em}\alpha\beta}F_{\alpha\beta}$
is the Hodge dual of the test function $F_{\alpha\beta}$, we have in effect ensured that the electromagnetic
field itself satisfies the Maxwell equation $\epsilon^{\alpha\lambda\mu\nu}\partial_\lambda F_{\mu\nu}=0$
in all states (in this last equation \underline{only}, $F_{\mu\nu}$ is what we might call the electromagnetic
field itself, which we can use here because there are no fluctuations whatsoever in the relevant degrees
of freedom, for all states that can be constructed by the application of creation operators to the vacuum
state, but care is required).

The Lie random field models of Section \ref{Schwartz} are completely determined by $\tilde G(u)$, with the
kernel of $f\mapsto f^\gimel$, $\widetilde{f^\gimel}(u)=\tilde G(u)\tilde f(u)$, determining
the kernel of $X\mapsto(X;X)$.
Linear constraints on observables and conserved currents make their appearance in this model as straightforward
projective properties of $\tilde G(u)$.
For nonlinear constraints and conserved currents, however, I am at present unable to offer a well-formulated
conjecture relating constraints and conserved currents to detailed properties of $f\mapsto f^\gimel$.

Although it appears that Lie random fields can be extended to include Fermionic algebraic elements, the standard
anti-commutation relations for spin-$\frac{1}{2}$ fields are not gauge-invariant, and I have not been
able to construct any that are.
It appears, therefore --- assuming that we take gauge invariance to have similar significance in the theoretical
setting of Lie random fields to its significance in Lagrangian and Hamiltonian approaches --- that we may
have to work with conserved observable currents and other constraints between observable fields, as
is commonly advocated in formal approaches to quantum field theory\cite{Fredenhagen}.
There are many alternatives to consider, however, and the constraints need to be articulated in terms
appropriate to Lie random field models in a way that might reproduce the phenomenology of quantum
electrodynamics, the electroweak model, or the standard model of particle physics.

Even if a Lie random field approach to fundamental physics ultimately fails, I have found that understanding this
kind of model relative to quantum field theory has greatly illuminated the latter, and Lie random fields may be
of use as practical models wherever classical stochastic methods have already been useful or would be useful
if quantum fluctuations were explicitly modeled.

\section{Lie random fields on cotangent manifolds}\label{TsM}
A natural extension of the scalar Lie random field on Minkowski space $M$ is a scalar Lie random field on the
cotangent manifold $T^*M$, for which we take the test functions to be maps
$f:T^*M\rightarrow\CC;(x,p)\mapsto f(x,p)$.
For some simple choices of detailed structure, this construction admits a classical relativistic gas
interpretation, but a wide range of detailed Lorentz invariant structures is possible for a Lie random field
on the cotangent manifold.
We will construct Lie random fields on the cotangent manifold for which we can take $f(x,p)$ to
be relatively unrestricted as a function of $p$ at a point $x$, while it is most convenient to require that
$f(x,p)$ should satisfy the Schwartz space requirements of smoothness and faster than polynomial
decrease as a function of $x$.

We will write $\tilde f(u,p)$ for the Fourier transform of $f(x,p)$ only in the $x$ variable, and we will
introduce a function $f^\gimel(x,p)$ for scalar test functions on the cotangent manifold,
$\widetilde{f^\gimel}(u,p)=\tilde G(u,p)\tilde f(u,p)$, for which
\begin{eqnarray}
   \label{TMeq1}
     (f;g)&=&\quad\int \widetilde{f^\gimel}^*(u,p)\widetilde{g^\gimel}(u,p)\Intd^4u\Intd^4p,\\
   \label{TMeq2}
     \widetilde{\xi^\gimel}(f;g)(s,p)&=&\lambda \int\delta^4(s+u-v)\widetilde{f^\gimel}^*(u,p)\widetilde{g^\gimel}(v,p)
                                         \Intd^4u\Intd^4v.
\end{eqnarray}
This constructs a Lie random field on the cotangent manifold if
$\tilde G^*(\frac{z-s}{2},p)\tilde G(\frac{z+s}{2},p)$ is invariant under $z\mapsto -z$ for all $p$ and $s$.
An instructive elementary example that satisfies this requirement is
\begin{equation}
  \tilde G(u,p)=\theta(u^2-m^2)\theta(p^2)\theta(p_0)
                           \exp{\left(-\;\frac{\textstyle(2\frac{(u\cdot p)^2}{u\cdot u}-p\cdot p)}{2m^2}\right)},
\end{equation}
which requires that $p$ must be small, in a Riemannian metric sense relative to the time-like direction of $u$,
for there to be a large contribution to inner products of test functions.
The particular choice of $\tilde G(u,p)$ given here ensures that there is a nonzero probability only if
$p$ is a forward-pointing 4-momentum vector, which allows us to interpret \emph{this} choice as somewhat
like a classical relativistic gas.

Moments of the probability density for $\hat\phi_f$ for this choice of $\tilde G(u,p)$ are finite even for
a test function such as $F(u)V_\mu p^\mu$ for any 4-vector $V_\mu$ and Schwartz space function $F(u)$
because of the ``coupling'' between $u$ and $p$.
Generally, we can use a test function $F(u)T(u,p)$ for any multinomial $T(u,p)$ in the 4-vectors $u_\mu$ and
$p^\mu$, including the most immediate examples of a scalar test function $F(u)$ and a set of test functions
indexed by a bivector, $F(u)B_{[\mu\nu]}u^{[\mu} p^{\nu]}$.
This type of model allows us to model measurements associated with a whole class of tensor-valued Lie random
fields on Minkowski space as measurements of a single scalar Lie random field on the cotangent manifold, which
would represent a significant unification if there are any physical systems for which it is possible.

We can also introduce a less convergent map $f\mapsto f^\gimel$, for which, for example,
\begin{equation}
  \tilde G(u,p)=\frac{\theta(u^2-m^2)\theta(p^2)\theta(p_0)}
                           {\left(m^4+2(u\cdot p)^2-(u\cdot u)(p\cdot p)\right)^{r/2}},
\end{equation}
which allows us to use a test function $F(u)T(u,p)$ for any multinomial $T(u,p)$ in the 4-vectors $u_\mu$ and
$p^\mu$ \emph{provided} the degree of $T(u,p)$ in $p$ is less than $r-2$.
For multinomials that exceed this degree, this ansatz does not give a test function that results in well-defined
measurements, so we can consider this map $f\mapsto f^\gimel$ to allow us to model a scalar field, a vector field,
and tensor fields that have degree less than $r-2$ in $p$, but there are also other fluctuations present that do
not allow test functions to be constructed using a tensor multinomial ansatz.
In general, the map $f\mapsto f^\gimel$ defines what test functions give well-defined measurements, but, except
for the specification of the kernel, there is an element of conventionality to our choice of $f\mapsto f^\gimel$.
Much more complicated functions $\tilde G(u,p)$ are of course possible, which presumably should be
classified by symmetries of their kernels.
A more general ansatz uses an arbitrary commutative and associative product for the cotangent manifold
coordinate (the most elementary example of which would be convolution instead of the straightforward
multiplication used in Equations (\ref{TMeq1}) and (\ref{TMeq2})\thinspace), which can be implemented in
terms of an integral transform $\widetilde{f^\gimel}(u,p)=\int\tilde G(u,p,q)\tilde f(u,q)\Intd^4q$, with
Equations (\ref{TMeq1}) and (\ref{TMeq2}) otherwise unchanged.
This constructs a Lie random field on the cotangent manifold if
$\tilde G^*(\frac{z-s}{2},p,q)\tilde G(\frac{z+s}{2},p,q')$ is invariant under $z\mapsto -z$ for all
$p$, $q$, $q'$, and $s$.

\section{Discussion}
We have proved the conjecture posed in \cite{MorganLF} that the vacuum vector allows us to construct a state
over a Lie random field, and hence we can construct a Hilbert space using the GNS construction.
We have also revisited the construction of Lie random fields on Minkowski space and introduced a construction
of Lie random fields on the cotangent space of Minkowski space, which it is hoped may allow a novel form
of unification of tensor-valued Lie random fields on Minkowski space.

The constraints of microcausality and of positive semi-definiteness of the Hamiltonian are so
stringent that after over 50 years only free quantum fields are known for the Minkowski space case,
suggesting that one or both constraints might profitably be weakened or removed.
The microcausality constraint is empirically quite well supported because of its close and much discussed
relationship to the impossibility of sending messages faster than light.
In contrast, that the Hamiltonian must be positive semi-definite is a mathematically forceful constraint
that is barely discussed in the literature, and we can straightforwardly construct a large class of what
otherwise appear to be interesting Lie random fields if we relax it.

It has also been understood that the violation of Bell inequalities prevents all classical models,
however random field models are capable of modeling the violation of Bell inequalities in exact parallel
with quantum field models.
Measurement can be modeled by nonlocal projection operators such as $a_g^\dagger\Vacuum\BraVacuum a_g$,
exactly as is commonly the practice in quantum optics, and adopting a pragmatic restriction to positive
frequencies, instead of modeling measurement by local operators such as $\hat\phi_f$.
Such nonlocal projection operators, using only positive frequency test functions satisfy exactly the same
noncommutative algebraic structure in a free random field model as they satisfy in a free quantum field
model, effectively as a mathematics appropriate for classical signal processing.
In the free random field case, the restriction to positive frequency is implemented by the modeler using
only positive frequency test functions, which we know from quantum field theory is empirically adequate,
instead of the restriction being enforced algebraically.
In a more general theoretical setting, it is in principle possible to model the violation of Bell
inequalities using random fields by virtue of the analysis in \cite{MorganBellRF}.
Note that the measurement theory associated with classical random fields is significantly different from
the measurement theory associated with quantum fields\cite[Appendix A]{MorganLF}.

It should be emphasized that we have not entirely avoided renormalization, but we have reformulated the
way in which characteristic length scales of an experimental apparatus are expressed in the mathematics.
Test functions used in a model of the state preparation and measurement parts of an experimental
apparatus define numerous length scales that are significant to the operation of the experiment.
Significantly, a test function characterization of the length scales that are most important for
understanding an experiment is better defined and more detailed than the single renormalization
mass scale that is introduced, in a relatively ad-hoc way, when using renormalization group methods
to characterize the entire operation of an experimental apparatus.
I cannot yet identify what properties of the Lie random field structure correspond to renormalizable
fields, but renormalizable fields are generally understood to be invariant whatever sequences of
functions we use when we approach a delta function limit.
It seems possible that such a criterion may characterize a tightly limited class of renormalizable
Lie random fields, however Lie random fields are relatively indifferent to such a criterion, and the
principle that we should only use renormalizable theories for fundamental physics is generally no
longer considered absolute.

There are two other technical issues that deserve mention.
We have not introduced a norm on the Lie random field algebras because the operators are unbounded, following
the practice of Wightman fields and of most particle physics.
We have not, in other words, engaged with the Haag-Kastler axiomatic approach, which posits the
existence of C${}^*$-algebras of bounded observables that satisfy the Haag-Kastler axioms.
Secondly, the formalism described here is manifestly Lorentz and translation covariant, but we have not
engaged with the axiomatic insistence that there must be a continuous action of the Poincar\'e group on the
Hilbert spaces we construct.
This paper, in contrast, would allow us to take a few test functions, freely generate a Hilbert space
using only those functions, then consider what measurements and state preparations can be modeled in
this limited regime.
Given a manifestly Lorentz and translation invariant formalism, we can construct an infinite set of test
functions, leading to a representation of the Poincar\'e group, and many of the resources of mathematical
analysis in consequence, but practical engineering often does not require the mathematics of infinite,
complete sets of test functions.

I am grateful to Marco Frasca for comments that redirected my thinking on positive semi-definiteness of
the Hamiltonian and to Ken Wharton for comments on a draft of the paper.

\appendix
\section{Positive semi-definiteness of the Hamiltonian: a discussion}\label{PositiveHamiltonian}
The concreteness of a Lie field approach results in a concentrated focus on positive semi-definiteness
of the Hamiltonian as constantly preventing the construction of what otherwise appear to be interesting
mathematical objects.

We assume here that we wish to preserve the associativity of the algebra of creation and annihilation
operators, the commutativity of creation operators and annihilation operators separately, $[a_X,a_Y]=0$ and
$[a_X^\dagger,a_Y^\dagger]=0$, and a $\star$-algebraic structure.
These are already fairly tight constraints, to which we would like to add at least two further axioms,
microcausality and positive semi-definiteness of the Hamiltonian on the vacuum sector.
It is apparent, however, that all these axioms together more-or-less over-constrain the algebraic structure,
in that, except for specific interacting quantum field models in low dimension, the only known quantum fields
are the free quantum fields\cite{Fredenhagen}.

Of the two additional axioms, microcausality seems more secure because of its relatively direct empirical
relationship to the impossibility of sending messages faster than light (although this does not dictate
that the field commutation relations must be nontrivial at time-like separation).
Positive semi-definiteness of the Hamiltonian, in contrast, although it is a centerpiece of both the
Wightman axioms and the Haag-Kastler axioms, is generally given only a weak theoretical justification that
it ensures ``stability'' of the vacuum: if there were lower energy states than the vacuum, surely the
vacuum would decay into them.
Against this reading of why we need positive semi-definiteness of the Hamiltonian in the vacuum sector,
thermal sector representations have an energy spectrum that is unbounded below, but we take the maximally
symmetric thermal state to be more thermodynamically stable than other states in a given thermal sector.
Positive semi-definiteness of the Hamiltonian is also central to algebraic approaches to quantum field theory
because it ensures various analyticity properties that are relied on heavily in proofs of various theorems,
however there are interesting theorems that can be proved in thermal sectors of a quantum field theory.

A Lie random field is essentially a classical field-theoretic model, but a relatively sophisticated
formalism is required to model Lorentz invariant vacuum states and thermal equilibrium states carefully.
Classical states that have nontrivial fluctuations have classical energy infinitely greater than states
that have no fluctuations, but may nonetheless be \emph{thermodynamically} stable if they are translation
invariant.
Positive semi-definiteness of the Hamiltonian looks rather different in the classical context of a Lie
random field than it looks for a quantum field, despite extensive algebraic and geometrical similarities.
Both positive and negative frequencies are essential to ensure that the classical trivial commutation relation
$[\hat\phi_f,\hat\phi_g]=0$ is satisfied for Lie random fields; the positive or negative frequency of a
mode of a Lie random field is not directly related to the infinitely positive classical energy of a
mode of a field that translation invariantly fills all of Minkowski space.

Finally, Lie random fields are constructed as a block-world formalism.
There is no question of ``stability'' in the conventional sense; adding a positive or a negative frequency
component to a given state just changes the results of measurement in different ways.
There is a block-world sense in which there is no evolution of a state, even though a unitary operator that
models active time translations can be constructed\cite[\S IV]{MorganLF}, because a state fixes what the
results of measurements would be for all time.
Energy arguments are less significant in a Lie random field formalism insofar as the Lie random field formalism is
treated more as a block-world model than is conventional perturbative quantum field theory.
Note that this is not an ontological claim that the world is a 4-dimensional block-world, only a description
of the mathematical components of this kind of model.


\begin{thebibliography}{99}
\bibitem{GreenbergA}
  {O. W. Greenberg, Ann. Phys. 16, 158 (1961).}

\bibitem{Lowenstein}
  {J. H. Lowenstein, Commun. Math. Phys. 6, 49 (1967).}

\bibitem{GreenbergB}
  {O. W. Greenberg, Commun. Math. Phys. 9, 13 (1968).}

\bibitem{Baumann}
  {K. Baumann, Commun. Math. Phys. 47, 69 (1976).}

\bibitem{MorganLF}
  {P. Morgan, J. Math. Phys. 48, 122302 (2007); arxiv:0704.3420 [quant-ph].}

\bibitem{delaPena}
  {L. de la Pen?a and A. M. Cetto, The quantum dice: an introduction to stochastic electrodynamics
     (Kluwer, Dordrecht, 1996).}

\bibitem{MorganBellRF}
  {P. Morgan, J. Phys. A 39, 7441 (2006); arxiv:cond-mat/0403692.}

\bibitem{Cohen}
  {L. Cohen, Proc. IEEE 77, 941 (1989).}

\bibitem{MorganQF}
  {P. Morgan, Phys. Lett. A 338, 8 (2005); arxiv:quant-ph/0411156.}

\bibitem{Haag}
  {R. Haag, Local Quantum Physics, 2nd Edition (Springer-Verlag, Berlin, 1996).}

\bibitem{MenikoffSharp}
  {R. Menikoff and D. H. Sharp, J. Math. Phys. 18 471 (1977).}

\bibitem{Fredenhagen}
  {K. Fredenhagen, K.-H. Rehren, and E. Seiler, Lect. Notes Phys. 721, 61 (2007); arxiv:hep-th/0603155.}

\end{thebibliography}
\end{document}